\documentclass[prb,twocolumn,showpacs,preprintnumbers,amsmath,amssymb]{revtex4}

\usepackage{graphicx,natbib,amssymb}
\usepackage{dcolumn}
\usepackage{bm}

\begin{document}


\title{Dependence of the structural and physical properties of Tl$_{1-y}$Fe$_{2-z}$(Se$_{1-x}$S$_{x}$)$_{2}$ with isovalent substitution of Se by S: decrease of $T_{N\acute{e}el}$ with S content}

\author{P. Toulemonde}
\altaffiliation {Corresponding author. Fax: 00334 7688 1038,
e-mail: pierre.toulemonde@grenoble.cnrs.fr}
\author{D. Santos-Cottin}
\author{Ch. Lepoittevin}
\author{P. Strobel}
\author{J. Marcus}
\affiliation{%
Institut N\'{e}el, CNRS and Universit\'{e} Joseph Fourier, 25
avenue des Martyrs, BP 166, F-38042 Grenoble cedex 9, France.
}%

\date{\today}

\begin{abstract}

The effect of selenium substitution by sulfur or tellurium in the
Tl$_{1-y}$Fe$_{2-z}$Se$_{2}$ antiferromagnet was studied by x-ray
and electron diffraction, magnetization and transport
measurements. Tl$_{0.8}$Fe$_{1.5}$(Se$_{1-x}$X$_{x}$)$_{2}$ (nominal composition) solid solutions were synthesized in the full x range (0~$\le x\le$~1) for X~=~S and up to x~=~0.5 for X~=~Te, using the sealed tube technique. No
superconductivity was found down to 4.2K in the case of sulfur
despite the fact that the optimal crystallographic parameters, determined
by Rietveld refinements, are reached in the series (i.e. the
Fe-(Se,S) interplane height and (Se,S)-Fe-(Se,S) angle for which
the critical superconducting transition T$_{c}$ is usually maximal
in pnictides). Quasi~full Tl site ($y\sim0.05$) compared to significant alkaline deficiency ($y=0.2-0.3$) in analogous A$_{1-y}$Fe$_{2-z}$Se$_{2}$ (A~=~K,~Rb,~Cs), and the resulting differences in iron valency, density of states and doping, are suggested to explain this absence of superconductivity.
Compounds substituted with tellurium, at least up
to x=0.25, show superconducting transitions but probably due to
tetragonal Fe(Se,Te) impurity phase. Transmission electron
microscopy confirmed the existence of ordered iron vacancies
network in the samples from the Tl$_{0.8}$Fe$_{1.5}$(Se$_{1-x}$S$_{x}$)$_{2}$ series in the form of the tetragonal $\sqrt{5}$ a $\times \sqrt{5}$ a $\times$ c superstructure (\textit{I4/m}) (mixed with the orthorhombic $\sqrt{2}$ a $\times 2\sqrt{2}$ a $\times$ c form (\textit{Ibam}) if the iron vacancies level is increased). The N\'{e}el temperature (T$_{N}$) indicating the onset of antiferromagnetism order in the $\sqrt{5}$ a $\times \sqrt{5}$ a $\times$ c supercell decreases from 450K in the selenide (x=0) to 330K in the sulfide (x=1).
 We finally demonstrate a direct linear relationship between $T_{N\acute{e}el}$ and the Fe-(Se,S) bond length (or Fe-(Se,S) height).

\end{abstract}



\maketitle

\section{Introduction}
After the discovery of superconductivity in iron-based
superconductors, i.e. in pnictides and chalcogenides, numerous
families were found, at least five families for arsenides with
superconducting transition up to T$_{c}$~=~55~K. In chalogenides,
superconductivity was first found in the \textquotedblleft
11\textquotedblright~family (Fe$_{1+y}$(Te$_{1-x}$Ch$_{x}$)) with
Ch=Se or S, and recently in a second family AFe$_{2-y}$Se$_{2}$
(\textquotedblleft A-122\textquotedblright~selenide) with
A=K~\cite{Guo1}, Rb, Cs (or Tl/Rb, Tl/Cs) showing T$_{c}$ around
30 K, i.e. close to the maximum value measured for FeSe under high
pressure~\cite{Mizuguchi,Garbarino1,Medvedev}. A related compound
is TlFe$_{2-y}$Se$_{2}$, which was first synthesized and studied
25 years ago by H\"{a}ggstr\"{o}m et al.~\cite{Haggstrom}. This compound
is antiferromagnetic with a high N\'{e}el temperature around
T$_{N}$=450K, i.e. in the range of T$_{N}$ values measured for
alkaline intercalated 122 selenides~\cite{Bao}.

In iron-based superconductors, superconductivity can be induced by
simple isovalent substitution of the pnictogen or chalcogen, for
example by substitution of As by P in LnFeAsO
(Ln~=~La,~Ce,~Pr,~Nd,~Sm...) (\textquotedblleft1111\textquotedblright)
or \textquotedblleft 122\textquotedblright~arsenides, or of Te by
Se/S in the Fe$_{1+y}$Te telluride. The present work follows the
same approach to search for superconductivity in thallium-122
selenide (\textquotedblleft~Tl-122\textquotedblright) by
substitution of selenium by sulfur. In addition, this substitution
may allow to approach the structural conditions where the highest
T$_{c}$'s are reached in this structural family, i.e. either a Fe-Se
bond length around 1.41~{\AA}~\cite{Okabe} (in the FeSe system
under high pressure) or Ch-Fe-Ch bond angles corresponding to a
regular FeCh$_{4}$ tetrahedron (in Fe-As systems)~\cite{Lee}.

During this study we became aware of an investigation of the
K$_{1-y}$Fe$_{2-z}$(Se$_{1-x}$S$_{x}$)$_{2}$ series by Lei et
al.~\cite{Lei}. In the potassium system, the x~=~0 end member is
already superconducting (T$_{c}$~=~33K) and superconductivity
disappears with increasing x(S). This may be related to (i) a
decrease in iron non-stoichiometry (i.e. the compound contains
less iron vacancies), (ii) an increase in FeCh$_{4}$ tetrahedron
distortion \cite{Lei}.

In this article we study the structural and physical trends vs. sulfur content
in the Tl$_{1-y}$Fe$_{2-z}$(Se$_{1-x}$S$_{x}$)$_{2}$ series.
Contrary to the alkaline-122 systems, the selenium-only and
end-member is known to present no superconductivity. We will show
in this article that all sulfur-substituted compositions remain
antiferromagnetic above room temperature, and that their N\'{e}el
temperature decreases linearly with decreasing Fe-Ch bond length.

\section{Experimental}

Tl$_{1-y}$Fe$_{2-z}$(Se$_{1-x}$S$_{x}$)$_{2}$ samples
(\textquotedblleft Tl-122(Se,S)\textquotedblright) were synthesized
using the sealed tube technique as reported elsewhere for
Fe$_{1+\delta}$(Te$_{1-x}$Se$_{x}$)~\cite{Klein,Noat}. Starting
materials were commercial Fe pieces (Alfa Aesar, 98{\%}), Tl
pieces (Alfa Aesar, 98{\%}), Se chips (Alfa Aesar, 98{\%}) and FeS
(Alfa, 99.9{\%}). Precursors with nominal composition
Tl$_{0.8}$Fe$_{1.5}$(Se$_{1-x}$S$_{x}$)$_{2}$ (i.e. z~=~0.5) were placed in an
alumina crucible which was introduced in a quartz tube and sealed
under vacuum. The heat treatment constituted of a first heating
ramp at 100\r{ }C/h up to 700\r{ }C followed by a plateau for
12~hours at this temperature; then the samples were slowly cooled at
5\r{ }C/h to 280\r{ }C and maintained at this temperature for 24h,
then furnace-cooled. We have also tried the substitution of Se by
Te (up to x=0.5) using Te pieces as precursor and a similar
temperature profile.

X-ray diffraction (XRD) patterns  were collected at room
temperature using a Bruker D8 powder diffractometer working in
Bragg-Brentano geometry at the wavelength $\lambda _{Cu,~K\alpha
1}$~=~1.54056~{\AA} from 2$\theta $~=~10 to 90\r{ } with a step of
0.032\r{ }.

Electron diffraction (ED) studies combined with quantitative
energy dispersive spectroscopy (EDS) micro-analysis were carried
out using a Philips CM 300 transmission electron microscope (TEM),
operating at 300 kV, equipped with a +/- 30\r{ } double tilt
sample holder. Specimens were prepared by crushing a small portion
of sample in an agate mortar containing ethanol in order to obtain
a powder with particles as thin as possible. Then a droplet of the
mixture was deposited on a copper grid with a holey carbon film,
in order to obtain an homogeneous particles distribution.

Transport measurements were carried out using the four point
contacts technique down to liquid helium temperature.
Magnetization of selected samples were measured at low (down to
4.2K) and high temperature (up to 600K) using a home-made
magnetometer with a resolution of 5.10$^{-6}$ A.m$^{2}$ and with
magnetic field up to 6T.

\section{Results and discussion}

\subsection{\label{sec:level1} Powder x-ray diffraction}

\begin{figure}
\begin{center}
\includegraphics[width=\linewidth]{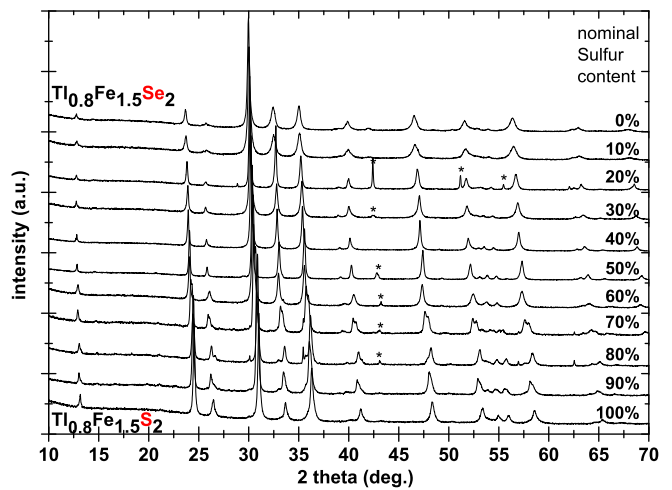}
\end{center}
\caption{XRD patterns ($\lambda $~=~1.5406~{\AA}) of
Tl$_{0.8}$Fe$_{1.5}$(Se$_{1-x}$S$_{x}$)$_{2}$ samples for $0 \le x
\le 1$. Nominal compositions (at. {\%} S) are indicated. Asterisk symbol indicate the main Bragg peaks of (non superconducting) hexagonal Fe(Se,S) impurity.}
\end{figure}

\begin{figure}
\begin{center}
\includegraphics[width=\linewidth]{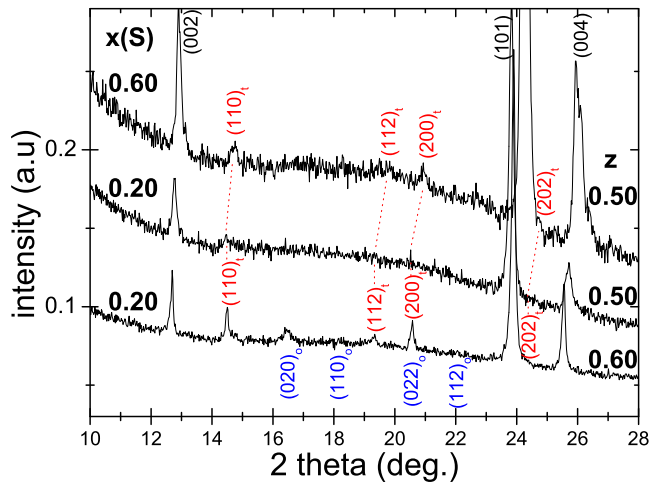}
\end{center}
\caption{Selected low 2-theta region of the XRD patterns of x=0.2 (with z=0.5 or z=0.6)
and x=0.6 Tl$_{0.8}$Fe$_{2-z}$(Se$_{1-x}$S$_{x}$)$_{2}$ samples
($\lambda $~=~1.5406~{\AA}). Supercell weak reflections related to
the tetragonal $\sqrt{5}$ a $\times \sqrt{5}$ a $\times$ c (Miller
indices labelled \textquotedblleft t\textquotedblright) or
orthorhombic $\sqrt{2}$ a $\times 2\sqrt{2}$ a $\times$ c
(labelled \textquotedblleft o\textquotedblright) superstructures
are indicated.}
\end{figure}

Figure 1 shows the powder x-ray diffraction (XRD) patterns of
polycrystalline Tl$_{0.8}$Fe$_{1.5}$(Se$_{1-x}$S$_{x}$)$_{2}$
samples (nominal composition) for sulfur contents from 0 to 100{\%}. Nearly all peaks
can be indexed in the tetragonal space group \textit{I4/mmm} found
for AEFe$_{2}$As$_{2}$ (AE~=~Ba,~Sr,~Ca) arsenides and used originally
by Guo et al.~\cite{Guo1} for their superconducting
KFe$_{2}$Se$_{2}$ selenide. This tetragonal structure is drawn in
the right part of fig.~3. Remaining unreacted (non superconducting) hexagonal Fe(S$_{1-x}$Se$_{x}$) is marked by an asterisk symbol. In addition we observe that a significant modification of the initial
nominal composition Tl:Fe:Ch~=~0.8:1.5:2 (for example an increase of iron content) induces the emergence of
the tetragonal Fe(Se$_{1-x}$S$_{x}$) secondary phase (XRD pattern not shown).

\begin{figure}
\begin{center}
\includegraphics[width=\linewidth]{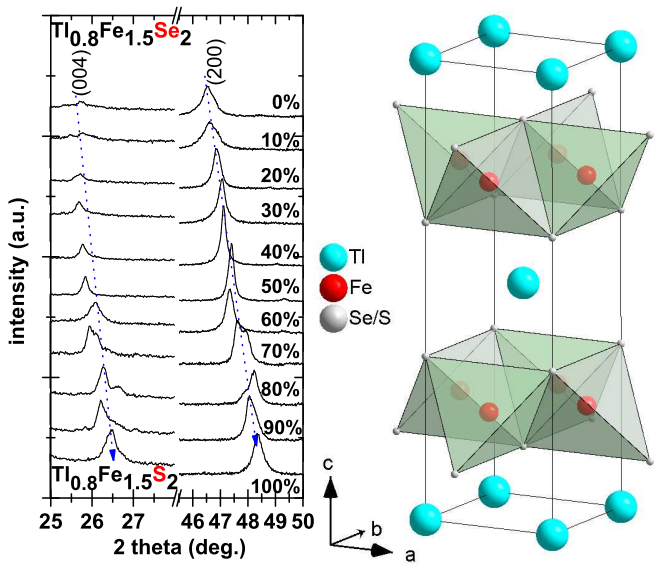}
\end{center}
\caption{Left: Selected 2-theta regions of the XRD patterns in the
Tl$_{0.8}$Fe$_{1.5}$(Se$_{1-x}$S$_{x}$)$_{2}$ series ($\lambda
$~=~1.5406~{\AA}) showing the 2 theta shift of (004) and (200)
reflections, corresponding to the shrinkage of the lattice with
x(S) increase. Right: Tl-122(Se,S) structure in the \textit{I4/mmm}
space group (i.e. with iron vacancies not ordered).}
\end{figure}

More interesting, for some sulfur contents, we clearly observe very
weak reflections (near the detection limit) at low angle. These
supplementary Bragg peaks can be indexed in supercells derived
from the original \textit{I4/mmm} lattice. If a and c represent
the subcell parameters, two superstructures were found in the
present study: a tetragonal $\sqrt{5}$~a~$\times~\sqrt{5}$~a~$\times$~c one (\textit{I4/m} space group) and an orthorhombic
$\sqrt{2}$~a~$\times~2\sqrt{2}$~a~$\times$~c one (\textit{Ibam}).
These superstructures are due to iron vacancy ordering observed for
z=0.4 and z=0.5 by Sabrowsky et al. 25 years ago in
TlFe$_{2-z}$S$_{2}$ sulfides~\cite{Sabrowsky}, and confirmed very
recently in the new alkaline-based selenides AFe$_{2-z}$Se$_{2}$
(A=K,Rb,Cs)~\cite{Pomjakushin,Song,Ye,Bao,Zavalij} but also in the
thallium-based one~\cite{Sales1}. An enlargement of the low
2-theta region of the XRD patterns for Tl$_{0.8}$Fe$_{1.5}$(Se$_{0.8}$S$_{0.2}$)$_{2}$ (i.e. z=0.5), Tl$_{0.8}$Fe$_{1.4}$(Se$_{0.8}$S$_{0.2}$)$_{2}$ (i.e. z=0.6) and Tl$_{0.8}$Fe$_{1.5}$(Se$_{0.4}$S$_{0.6}$)$_{2}$  is displayed figure 2. For x=0.6 (and z=0.5) the satellite peaks of the $\sqrt{5}$~a~$\times~\sqrt{5}$~a~$\times$~c supercell are clearly visible. For the sample with less sulfur, x=0.2 and the same nominal iron content (z=0.5) these satellite peaks are still present (the highest one being the (110) reflection)) but seem less intense. In contrast, when the level of iron vacancies is increased, i.e. for z=0.6 (keeping x=0.20), new small satellite peaks appear, in addition to those related to the $\sqrt{5}$~a~$\times~\sqrt{5}$~a~$\times$~c supercell, which can be indexed in the orthorhombic $\sqrt{2}$~a~$\times~2\sqrt{2}$~a~$\times$~c supercell. This result, confirmed by electron diffraction (see below), is in agreement with the previous work of Sabrowsky~\cite{Sabrowsky} who reported the orthorhombic supercell for low level of iron content in sulfides.

Despite the difficulty to detect the satellite reflections by conventional powder XRD, this XRD characterization suggests that all samples contain at least the $\sqrt{5}$~a~$\times~\sqrt{5}$~a~$\times$~c phase (mixed with the orthorhombic $\sqrt{2}$~a~$\times~2\sqrt{2}$~a~$\times$~c one if the nominal content of iron is decreased). To confirm this result we have performed an electron diffraction study of selected samples.
In addition, because supercell reflections are extremely weak (near the detection level of our diffraction setup), all the Rietveld refinements of our XRD patterns were performed in the average \textit{I4/mmm} space group.

Concerning the tellurium-substituted
Tl$_{1-y}$Fe$_{2-z}$(Se$_{1-x}$Te$_{x}$)$_{2}$, the synthesis of
phase pure samples is more difficult. Above x(Te)~=~0.2 the samples
contain secondary phases: the tetragonal form of
Fe(Te$_{1-x}$Se$_{x}$), and FeTe$_{2}$ and Tl$_{5}$Te$_{3}$
tellurides. We will not discuss these samples in detail in this
paper.

\begin{figure}
\begin{center}
\includegraphics[width=\linewidth]{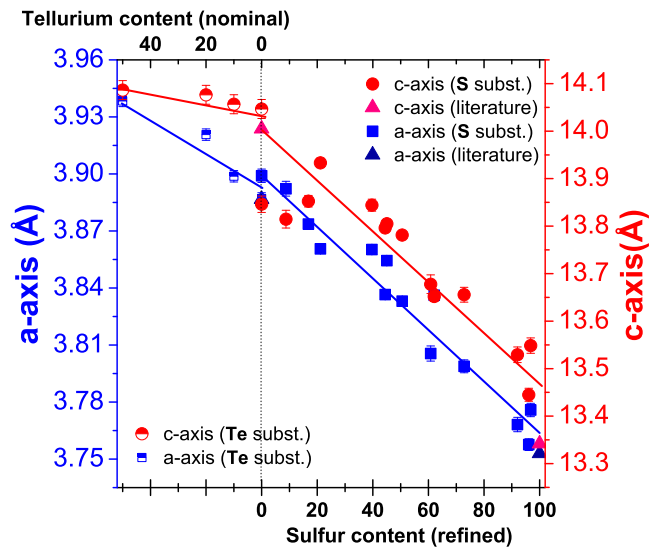}
\end{center}
\caption{Refined lattice parameters of the
Tl$_{1-y}$Fe$_{2-z}$(Se$_{1-x}$X$_{x}$)$_{2}$ samples with X=S or
Te (from Rietveld refinement of the XRD patterns) as a function of
S/Te substitution level. Values from literature are taken from
ref.~\cite{Haggstrom} for the pure selenide (x=0) and
ref.~\cite{Klepp} for the pure sulfide (x=1).}
\end{figure}

\begin{figure}
\begin{center}
\includegraphics[width=\linewidth]{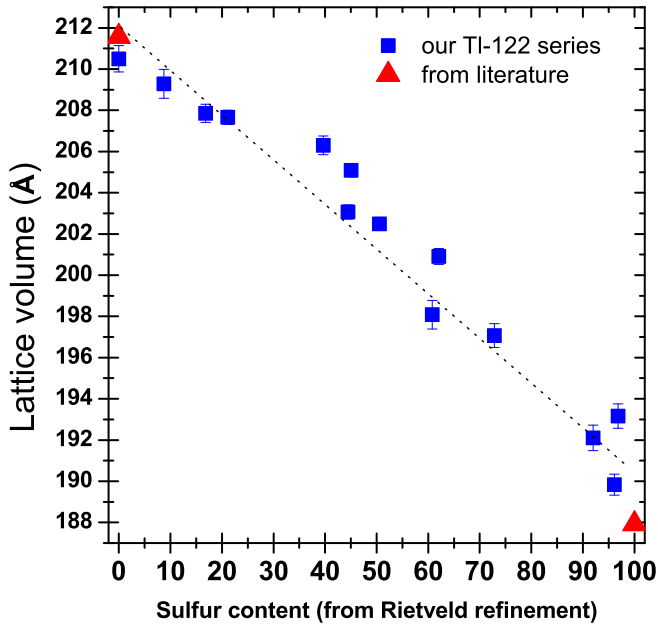}
\end{center}
\caption{Sulfur content dependence of the volume lattice in the
Tl$_{1-y}$Fe$_{2-z}$(Se$_{1-x}$S$_{x}$)$_{2}$ samples.}
\end{figure}

An enlargement of two regions of the XRD patterns around (004) and
(200) reflections of the Tl-122(Se,S) series shows a continuous shift to higher angle with
sulfur content (Figure~3). This corresponds to a decrease of both
lattice parameters of the Tl-122(Se,S) lattice with increasing x(S). The evolution of cell parameters and volume of the unit cell with sulfur content, determined from Rietveld refinement of XRD patterns are displayed respectively in figure~4 and~5. The decrease observed here (from a~{$\sim3.88$~\AA} and
c~{$\sim14$~\AA} for x(S)~=~0 to a~{$\sim3.75$~\AA} and c~{$\sim13.4$~\AA} for x(S)~=~1, i.e. a reduction of the unit cell volume of about 10{\%}, fig. 5) is similar in amplitude to that reported in the potassium K$_{1-y}$Fe$_{2-z}$(Se$_{1-x}$S$_{x}$)$_{2}$ series (see fig. 1 in ref.~\cite{Lei}). The lines (guide for eyes) in fig.~4 show that
our samples are in agreement with the expected values considering
a linear decrease between extremal x~=~0 and x~=~100{\%} compositions.
The small deviation from this linear trend (visible in the c-axis
variation for $x~<~20{\%}$ for example) is probably due to small
variations in iron and/or thallium contents between different
samples (see the trend shown in table~1 for refined values of Fe
and Tl site occupancy factors). This is also an indication that
the description in the average \textit{I4/mmm} space group is not
fully correct and that the real space group should be less
symmetric (i.e. taking into account ordered iron vacancies).
About the real stoichiometry of the samples, as we will see it later in part D (Rietveld refinements results), we found that the Tl site is nearly full despite the fact that the nominal composition corresponds to 20{\%} of deficiency (y~=~0.2). This is an important difference with alkaline based A-122 selenides where real alkaline deficiency is generally around 20-30{\%} in superconducting compounds~\cite{Pomjakushin,Zavalij}.

In contrast to the shrinkage of the unit cell observed in the sulfur substituted samples, tellurium substituted samples Tl$_{1-y}$Fe$_{2-z}$(Se$_{1-x}$Te$_{x}$)$_{2}$ show the expected increase of both c-axis, up to $c\sim14.09~\AA$ for x(Te)~=~0.5, and
a-axis, up to $a\sim3.94~\AA$ for x(Te)~=~0.5 (see left part of
fig.~4), considering the larger atomic radius of tellurium
compared to the selenium one. The rate of this increase is similar
to the one observed in the tetragonal Fe(Te$_{1-x}$Se$_{x}$)
series and seems more important for the a-axis than for the
c-axis; this trend is opposite to that observed in alkaline-based
Rb$_{0.8}$Fe$_{2-y}$(Se$_{1-x}$Te$_{x}$)$_{2}$ series where c-axis
increases faster than a-axis with Te content~\cite{Gu}.

\subsection{Electron diffraction in TEM}

x=0, 0.2 (with z~=~0.6) and 0.7 compositions were selected for detailed electron
diffraction studies.

\begin{figure}
\begin{center}
\includegraphics[width=0.8\linewidth]{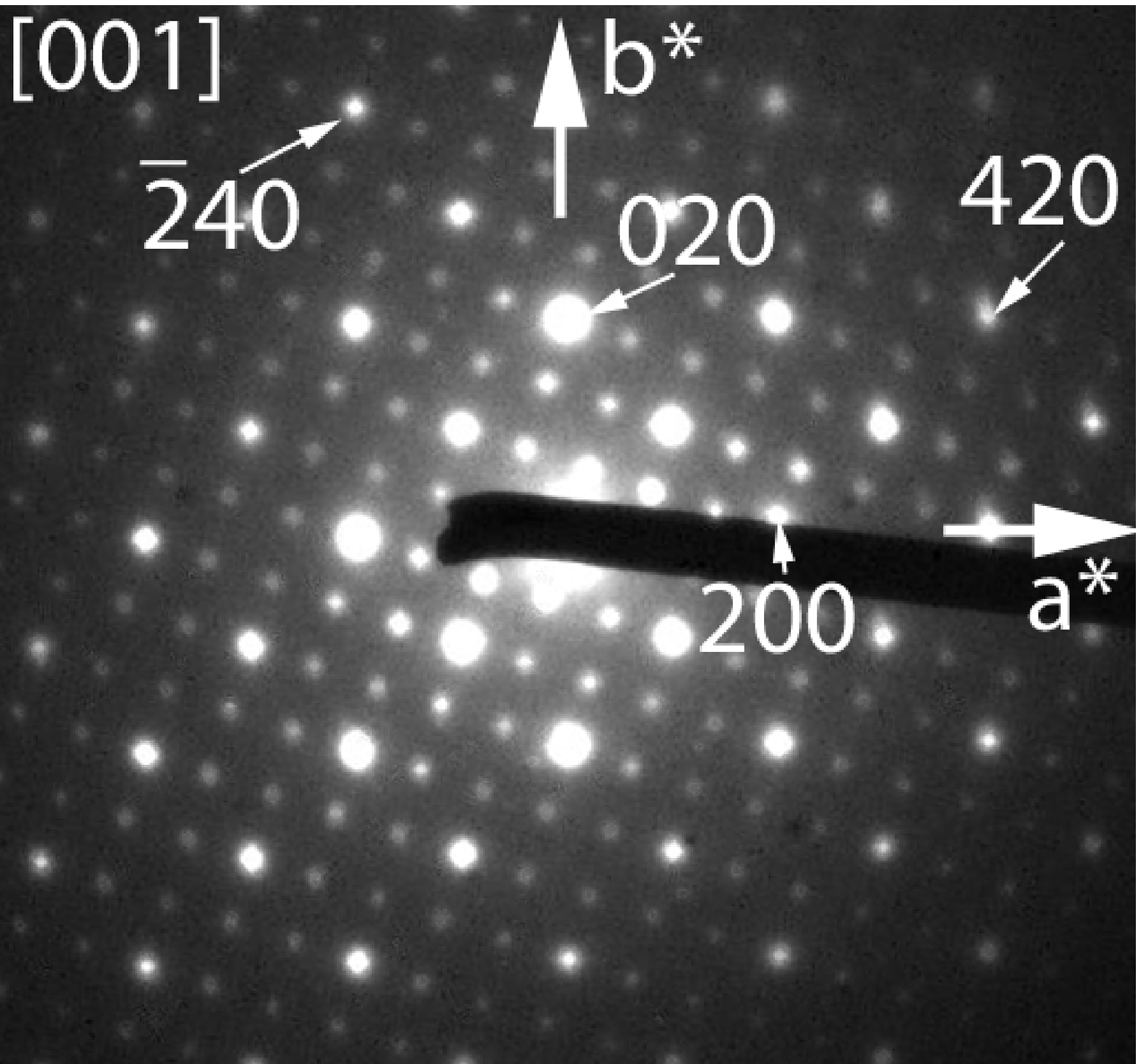}
\includegraphics[width=0.8\linewidth]{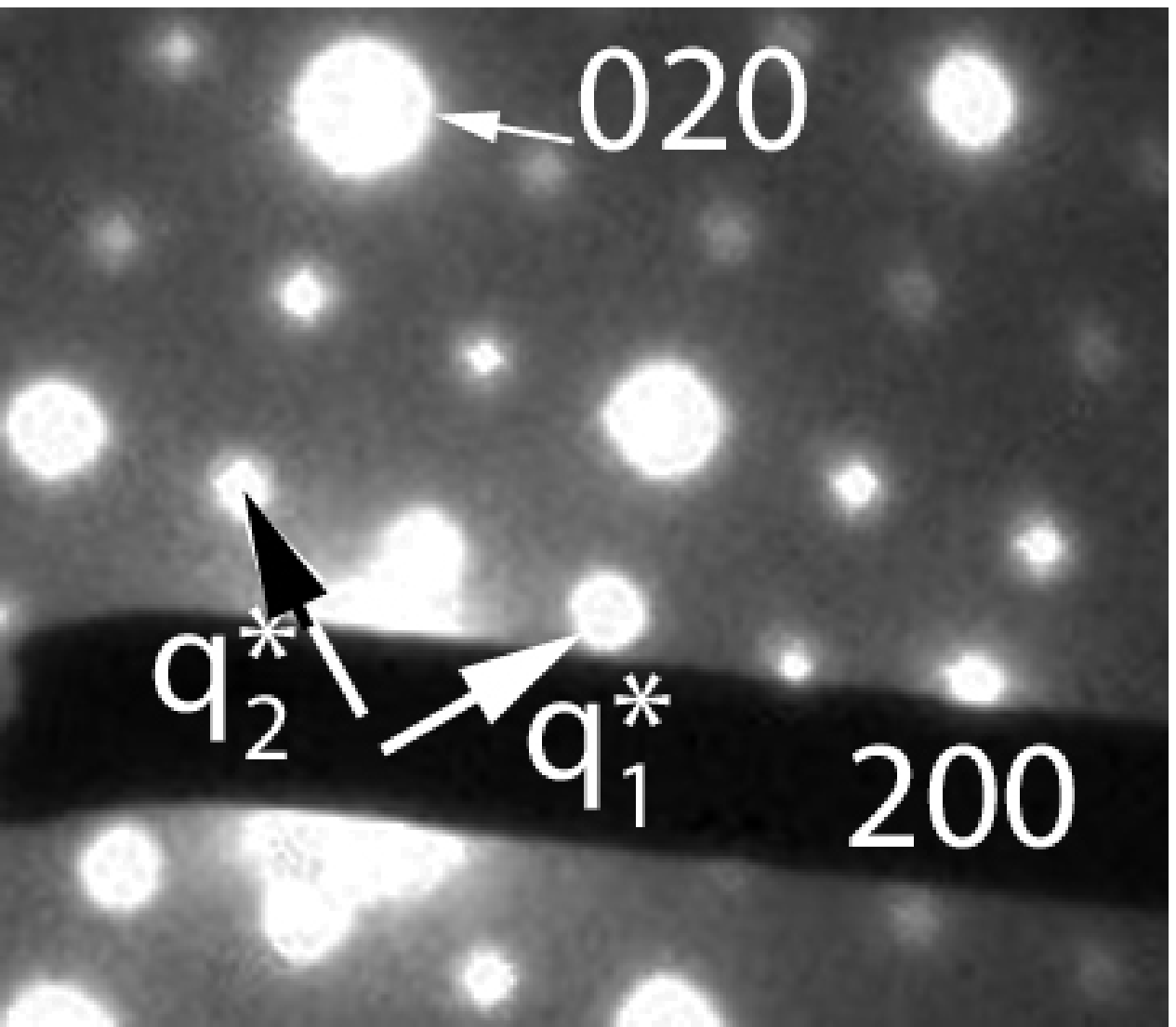}
\end{center}
\caption{(Top)[001] ED pattern of x=0 sample (Tl$_{1.0(1)}$Fe$_{1.7(1)}$Se$_{2}$
from EDS) indexed in a tetragonal sub-cell. (Bottom) The zoomed
area evidences the two directions of the modulation vector
associated to the satellite reflections.}
\end{figure}

For the pure selenide (x~=~0, z~=~0.5), the EDS
analysis carried out on around 50 crystallites shows an
homogeneous average cationic composition Tl$_{1.0(1)}$Fe$_{1.7(1)}$Se$_{2}$,
i.e. it contains a significant amount of iron vacancies. The
electron diffraction patterns recorded on different particles
present a body-centered tetragonal sub-cell with a~=~b~=~3.9~{\AA},
c~=~14~{\AA}. The extinctions observed are compatible with the
\textit{I4/mmm} space group. However, extra reflections, called
satellite reflections, can be observed on the [001] oriented basal
plane (figure 6). These extra spots are characteristic of a
modulated structure with a two-components modulation vector
$\vec{q^{*}}= \alpha \vec{a^{*}} + \beta \vec{b^{*}}$. According
to this ED pattern, there are several ways to define the
modulation vectors. We chose here two vectors in agreement with
the superstructure defined previously by Pomjakushin et
al.~\cite{Pomjakushin} for the Cs-based 122 selenide:
$\vec{q^{*}_{1}}$ and $\vec{q^{*}_{2}}$ lie along [210] and
[$\bar{1}$20] directions of the subcell, with an amplitude of 1/5,
leading to the values
$\vec{q^{*}_{1}}$~=~1/10(-2$\vec{a^{*}}$+4$\vec{b^{*}}$) and
$\vec{q^{*}_{2}}$~=~1/10(4$\vec{a^{*}}$+2$\vec{b^{*}}$). Bearing
in mind the commensurate nature of the modulation, the structure
can also be described in a supercell a~=~b~=~8.7~{\AA} (=$\sqrt{5}$
a), c~=~14~{\AA} (\textit{I4/m}). This result is in agreement
with neutron and x-ray powder and single crystal diffraction data
reported recently on Cs$_{y}$Fe$_{2-x}$Se$_{2}$
system~\cite{Pomjakushin} and with the electron diffraction study
of K$_{y}$Fe$_{2-x}$Se $_{2}$ showing evidence of a
$\sqrt{5}$ a $\times \sqrt{5}$ a $\times$ c supercell\cite{Wang}.

\begin{figure}
\begin{center}
\includegraphics[width=0.8\linewidth]{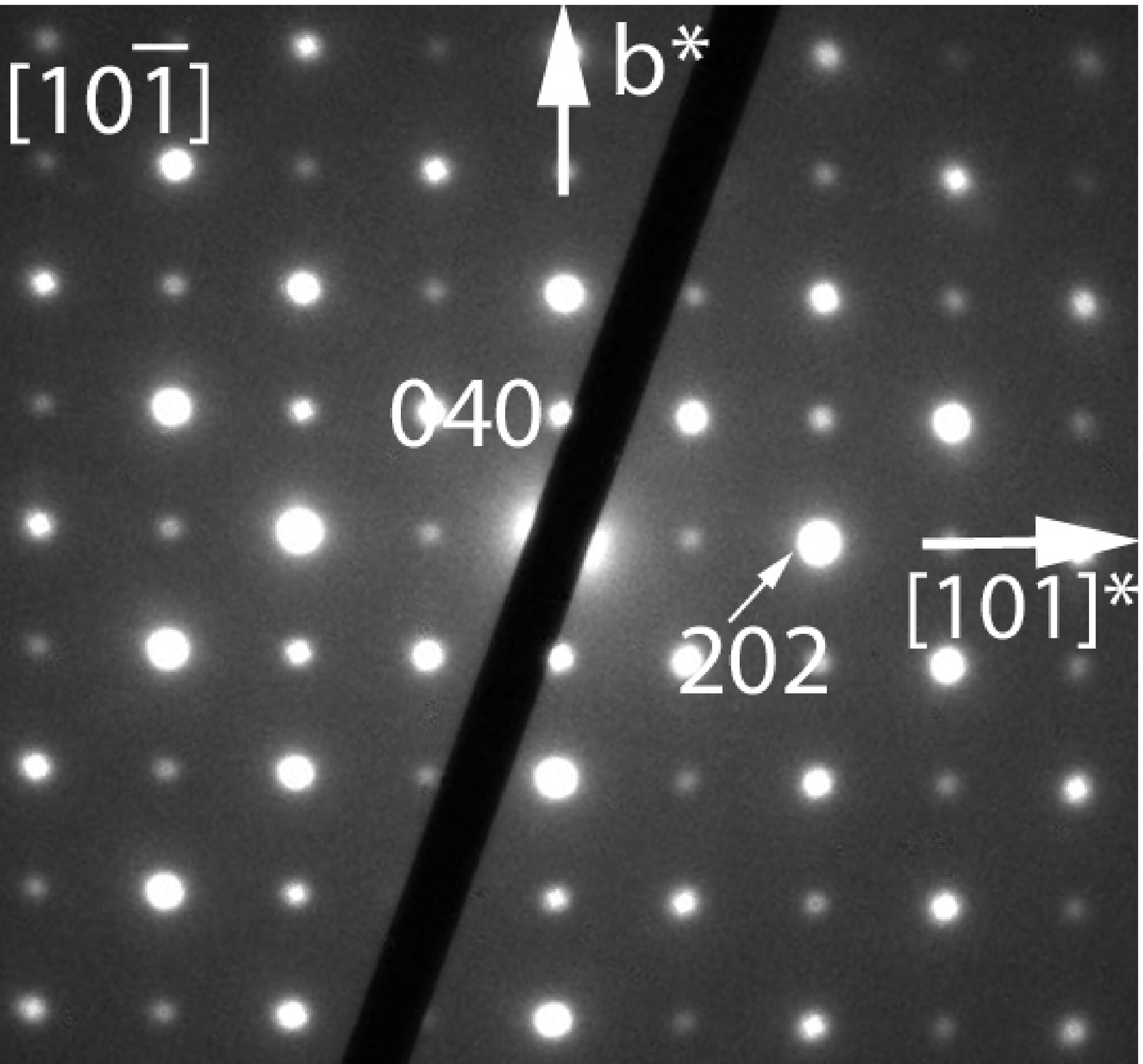}
\includegraphics[width=0.8\linewidth]{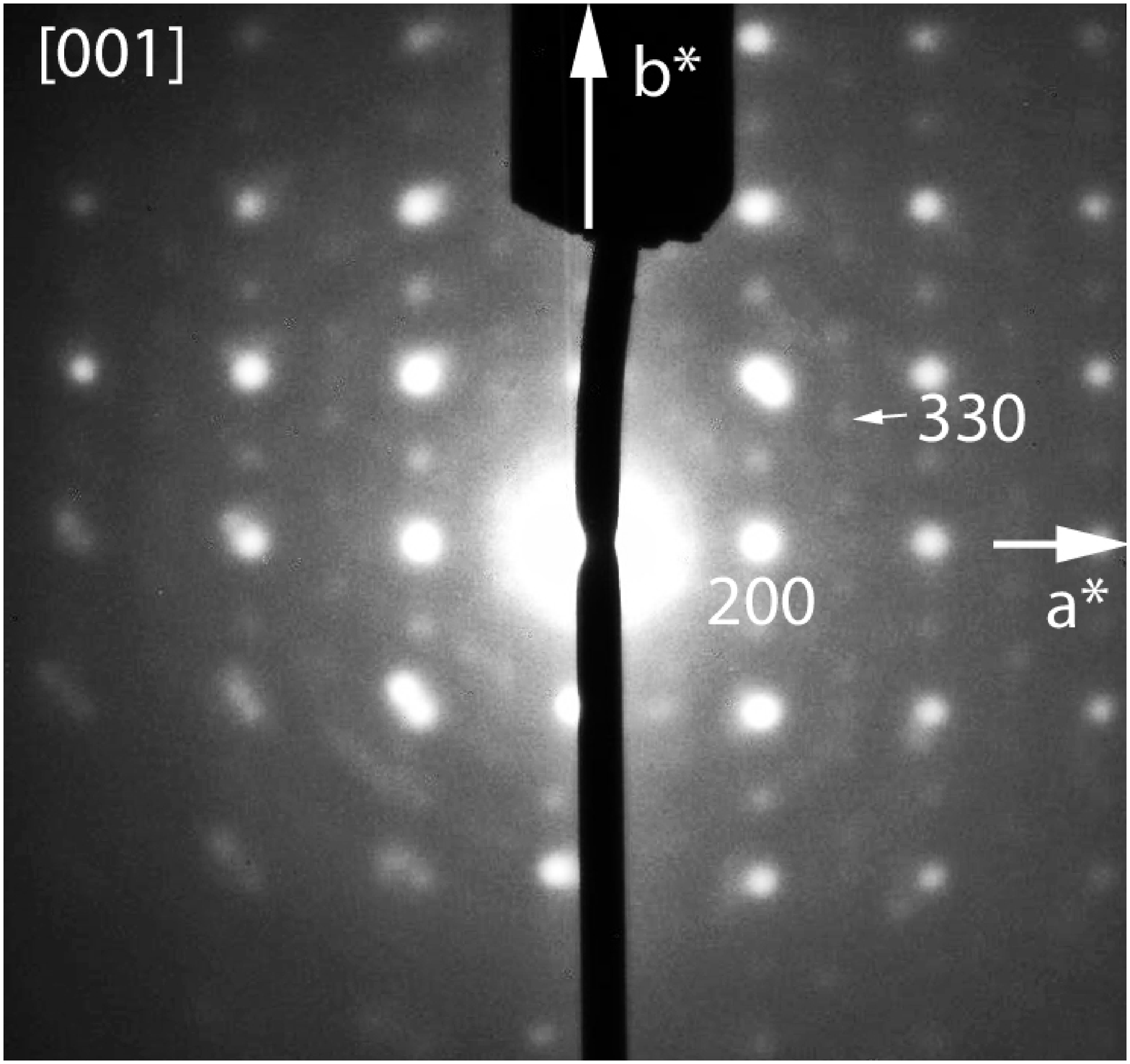}
\end{center}
\caption{[10$\bar{1}$] (top) and [001] (bottom) ED pattern of
x=0.2and z=0.6 sample (Tl$_{0.8(1)}$Fe$_{1.4(1)}$(Se$_{0.75(5)}$S$_{0.25(5)}$)$_{2}$
from EDS) indexed in an orthorhombic cell.}
\end{figure}

For the Se-rich composition (x~=~0.2) and higher nominal iron vacancy level (z~=~0.6 instead of z~=~0.5 in the series, see XRD pattern, fig.~2), the EDS analysis carried out on
numerous crystallites confirms a homogeneous cationic composition
Tl$_{0.8(1)}$Fe$_{1.4(1)}$Se$_{1.5(1)}$S$_{0.5(1)}$, not far from the nominal composition, i.e. with less iron than the previous sample. We note also that this sample
contains a lower content of thallium than the x=0 sample.
The reconstruction of the reciprocal space obtained by tilting
around the b* crystallographic axis led to an orthorhombic cell
with the parameters a~=~5.6~{\AA} ($\sqrt{2}$ a), b~=~11.3~{\AA}
(2~$\sqrt{2}$ a) and c~=~15~{\AA} (figure 7). The reflexions
conditions (hkl : h+k+l=2n, 0kl : k = 2n and h0l : h = 2n) are
compatible with space group \textit{Ibam} (n\r{ }72). Note that
the 101 and 303 reflections visible on the [10$\bar{1}$] ED
pattern are artefacts caused by the multiple diffraction ; upon
rotation around the [101] axis, these reflections indeed disappear
depending on the zone axis. We conclude that these crystallographic features are in
agreement with the indexation of extra weak peaks of the x=0.2, z=0.6 XRD
pattern (see fig.~2) , and also with the orthorhombic structure
obtained for the TlFe$_{1.5}$S$_{2}$ pure sulfide in
1980~\cite{Zabel}.

\begin{figure}
\begin{center}
\includegraphics[width=0.8\linewidth]{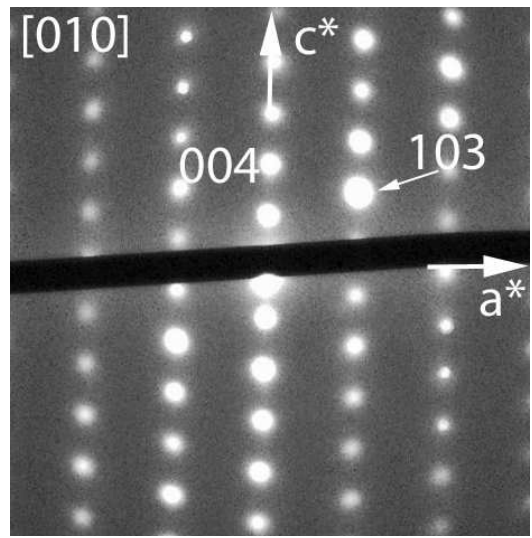}
\end{center}
\caption{[010] ED pattern of x=0.7 sample
(Tl$_{0.8(1)}$Fe$_{1.5(1)}$(Se$_{0.3(1)}$S$_{0.7(1)}$)$_{2}$ from EDS) indexed
in a tetragonal sub-cell.}
\end{figure}

For a S-rich system (x=0.7, z~=~0.5) of the Tl-122(Se,S) series, a very homogeneous cationic
composition close to the nominal formulation was revealed by EDS
analysis. Electron diffraction evidenced the tetragonal structure
as for x=0. Figure 8 exhibits an ED pattern recorded along the
[010] zone axis with the parameters $a\sim 3.9$~{\AA} and $c\sim
14$~{\AA}.

As a conclusion, this electron diffraction study of selected x(S) compositions confirm that the extra peaks shown by x-ray diffraction (at the limit of detection level) are those due to the superstructure which appear when the iron vacancies are ordered in their basal plane. It confirms that all investigated samples contain at least a major fraction crystallized in the iron ordered tetragonal $\sqrt{5}$ a $\times \sqrt{5}$ a $\times$ c superstructure (\textit{I4/m}), and another part in the orthorhombic $\sqrt{2}$ a $\times 2\sqrt{2}$ a $\times$ c superstructure (if the iron vacancy content z is increased), and/or a minor part with the iron disordered \textit{I4/mmm} lattice.

\subsection{\label{sec:level1} Electrical resistance and magnetization}

Figure 9 shows the typical magnetization curve M(T) and electrical
resistance R(T) for x=0.4 sample as a function of (high)
temperature. For this composition, we clearly see the onset of the AFM ordering around
390K on the M(T) curve. The appearance of the magnetic ordering
is also visible in the R(T) curve as an anomaly which is more or
less marked depending on the sulfur content. This magnetic
transition (at T$_{N}$) is preceded by a structural transition (at
T$_{S}$ slightly above T$_{N}$) which corresponds to the ordering
of iron vacancies (disordered at high T, i.e. corresponding to the
\textit{I4/mmm} description) in the tetragonal $\sqrt{5}$ a $\times \sqrt{5}$ a $\times$ c superstructure, as shown by our combined XRD and ED studies. A similar behavior was also reported in alkaline based selenides in previous neutron diffraction studies of K$_{0.8}$Fe$_{1.6}$Se$_{2}$~\cite{Bao}
(T$_{N}$=559K and T$_{S}$=578K) and Rb$_{0.8}$Fe$_{1.6}$Se$_{2}$
(T$_{N}$=502K and T$_{S}$=515K) or Cs$_{0.8}$Fe$_{1.6}$Se$_{2}$
(T$_{N}$=471K and T$_{S}$=500K)~\cite{Ye}. In thallium
phases T$_{N}$ and T$_{S}$ seem to be very close (fig.~9), and we
have used the anomaly in the transport measurements as a
determination of T$_{N}$, as usually made in analogous
alkaline-based AFe$_{2-y}$Se$_{2}$ selenides
(A=K,Rb,Cs)~\cite{Liu,Song}.

\begin{figure}
\begin{center}
\includegraphics[width=\linewidth]{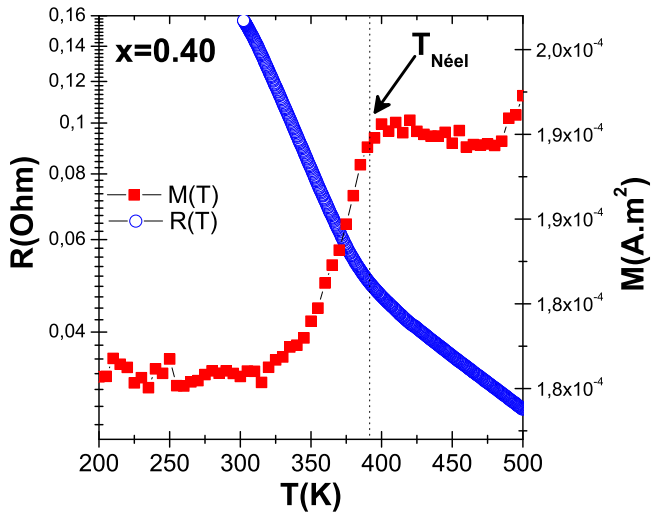}
\end{center}
\caption{Magnetization measured at 6T and electrical resistance
(in log scale) in the 200-500K range of the x=0.4 sample. In both
measurement the signature of the onset of the AFM order at T$_{N}$
is clearly visible.}
\end{figure}

The next figure (fig.~10) shows the electrical resistance (R(T))
behaviors (in log scale) of the two extremal composition x=0 and
x=1 in a larger temperature range, i.e. from 600K down to the
liquid helium temperature. All compositions showed a
semi-conducting behavior at low T, independently of the sulfur
content. No superconductivity has been found down to 4.2K. In the
particular case of the pure selenide (x=0), in the 4-300K range,
the R(T) curve of our polycrystalline sample shows two regimes
which intersect around T$_{2}$=120K, as observed previously by
Sales et al. at 100K on a TlFe$_{1.6}$Se$_{2}$ single
crystal~\cite{Sales1}. These authors have also evidenced another
transition temperature around T$_{1}$=150K, not visible in our
samples, based on their specific heat, magnetization and transport
measurements. And they have concluded very recently that this
particular behavior of TlFe$_{1.6}$Se$_{2}$ between T$_{1}$ and
T$_{2}$ was related to a sudden change of z position of 4 Fe spins
pointing down (along c-axis) and 4 Fe spins pointing up in the
\textquotedblleft block checkboard\textquotedblright AFM
structure, inducing a corrugation of the iron layer and a canting
of the Fe magnetic moment relatively to the c-axis (up to 27(3)\r{
} at 115K) for T$_{2}<T<T_{1}$~\cite{Cao}. We are unable with the
present data to discuss in details this point, but it seems that
T$_{2}$ decreases with the sulfur content introduced in the
lattice. It will be maybe interesting in the future to study if
this unusual magnetoelastic behavior seen in the selenide case for
T$_{2}<T<T_{1}$ persists also up to x(S)=1 and how it evolves.

\begin{figure}
\begin{center}
\includegraphics[width=\linewidth]{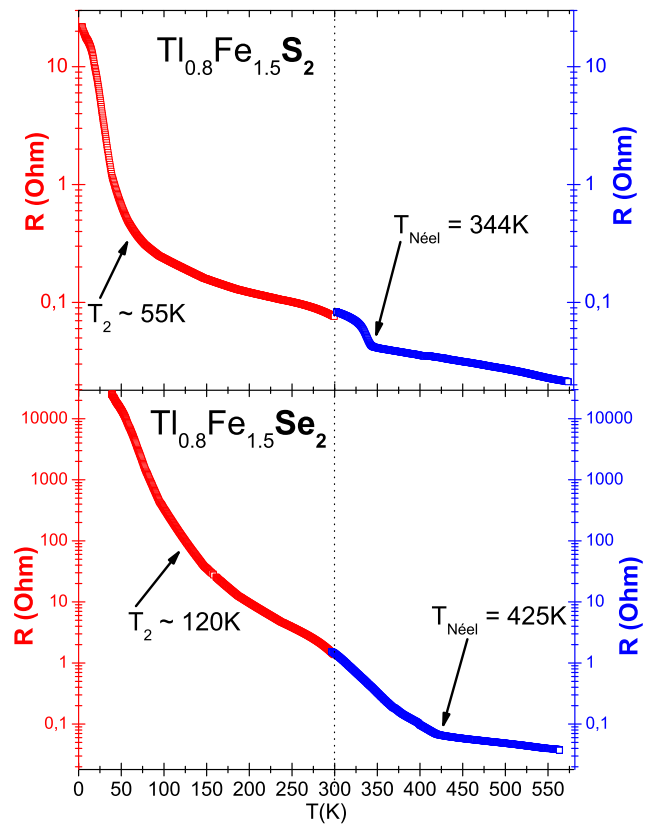}
\end{center}
\caption{Low and high temperature dependance of the electrical
resistance of typical x=0 and x=1.0 samples of  of the Tl$_{0.8}$Fe$_{1.5}$(Se$_{1-x}$S$_{x}$)$_{2}$ series emphasizing
the signature of the long range AFM structure in the ordered iron
vacancies network at T$_{N}$.}
\end{figure}

\begin{figure}
\begin{center}
\includegraphics[width=\linewidth]{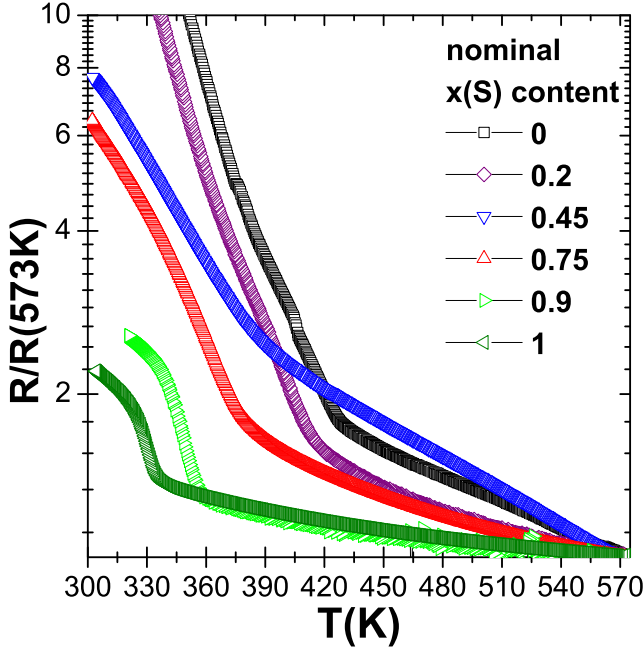}
\end{center}
\caption{High temperature dependance of the electrical
resistance of several Tl$_{0.8}$Fe$_{1.5}$(Se$_{1-x}$S$_{x}$)$_{2}$ samples with $0 \le x
\le 1$. Nominal sulfur x(S) content is indicated. The anomaly related to the long range AFM structure in the ordered iron
vacancies network at T$_{N}$ is clearly shown and decreases with x(S).}
\end{figure}

\begin{figure}
\begin{center}
\includegraphics[width=\linewidth]{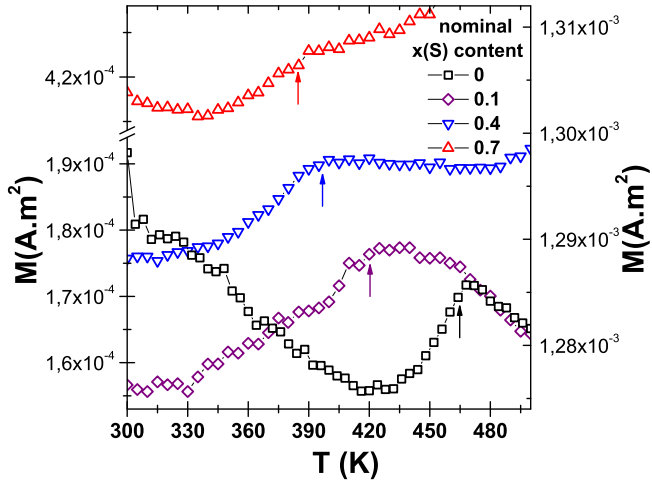}
\end{center}
\caption{High temperature dependance of the magnetization (measured at 6T) of several Tl$_{1-y}$Fe$_{2-z}$(Se$_{1-x}$S$_{x}$)$_{2}$ samples for nominal x(S)=0 (and nominal composition Tl$_{1}$Fe$_{1.8}$Se$_{2}$; right scale), 0.1, 0.4 and 0.7. Nominal S/Se ratio x(S) content is indicated. The onset of the long range AFM ordering in the ordered iron vacancies network at T$_{N}$ is marked by an arrow and is found to decrease with x(S).}
\end{figure}

The characteristic temperature T$_{N}$ extracted from the anomaly observed in the high temperature resistivity measurement, and associated with the antiferromagnetism ordering at low T, is clearly visible for all x(S) compositions (see curves normalized to the value measured at 573K, figure 11). It unambiguously decreases gradually in the Tl$_{0.8}$Fe$_{1.5}$(Se$_{1-x}$S$_{x}$)$_{2}$ series from T$_{N}$=425K for the selenide (x=0) to T$_{N}$=344K for the sulfide (x=1). We also note that for a fixed value of x(S), T$_{N}$ can also be slightly changed by tuning the iron content in the lattice (i.e. the iron vacancies content), via the change of the nominal ratios Tl:Fe:Se+S (T$_{N}$ is increased to 475K for example in Tl$_{1}$Fe$_{1.8}$Se$_{2}$, see figure 12). However, we did not not systematically explored this trend and focused our study on the series with a nominal ratio Tl:Fe equal to 0.8:1.5.
This transition from a paramagnetic state to an antiferromagnetic one at low temperature is less detectable in our magnetic measurements. This is particular true for high values of x(S), as shown in the curves of figure 12, because of a supplementary background arising from magnetic impurities contained in our samples (quasi undetectable by XRD). For this reason, as mentioned at the beginning of this paragraph, we have preferred the anomaly seen in the transport measurements for the determination of T$_{N}$.

And finally, transport measurements in the Tl$_{1-y}$Fe$_{2-z}$(Se$_{1-x}$Te$_{x}$)$_{2}$ Se-Te system showed
superconductivity (with zero resistivity) around T$_{c}$=15K (not
shown); the corresponding magnetization measurements gave a very
small diamagnetic shielding. We therefore conclude that this
superconductivity is probably due to the tetragonal
Fe(Te$_{1-x}$Se$_{x}$) secondary phase, in agreement with its
fraction estimated from x-ray diffraction. In addition, we note
that tellurium substitution in analogous superconducting alkaline-based
Rb$_{0.8}$Fe$_{2-y}$(Se$_{1-x}$Te$_{x}$)$_{2}$ induces a strong
reduction of superconductivity which disappears for
x(Te)=0.15~\cite{Gu}.

\subsection{Rietveld refinements in the average \textit{I4/mmm} space group}

\begin{table*}
\caption{\label{tab:table1} Refined structural parameters of x=0,
x=0.5 and x=1 compositions of
Tl$_{1-y}$Fe$_{2-z}$(Se$_{1-x}$S$_{x}$)$_{2}$ series using the
average \textit{I4/mmm} description (Ch=Se,S).}
\begin{ruledtabular}
\begin{tabular}{l|c|c|c|c|c|c}
compound & x=0 & x=0  at 250K & x=0 & x=0.5 & x=1 & x=1 \\
   & H\"{a}ggstr\"{o}m et al.~\cite{Haggstrom} & Cao et al.~\cite{Cao} & & & & Klepp and Boller ~\cite{Klepp} \\
   & (1986) & (2012) & & & & (1978) \\ \hline
 a-axis ({\AA}) & 3.8867(3) & 3.884(2)$^{\rm d}$ & 3.8870(1) & 3.8331(2) & 3.7572(3) & 3.755(1) \\
 c-axis ({\AA}) & 14.005(1) & 14.002(7) & 14.0401(3) & 13.781(1) & 13.443(2) &  13.35(1) \\
 n(Tl) & 1 & 1 & 0.98(2) & 0.90(2) & 0.92(2) & 1 \\
 n(Fe) & 1 & 0.795(5)$^{\rm e}$ & 0.66(2) & 0.71(1) & 0.70(1) & 1 \\
 n(S) & - & - & -& 0.52(2) & 1 & 1 \\
 z(Ch) & 0.357$^{\rm a}$ & 0.3575(2)$^{\rm f}$ & 0.3530(2) & 0.3541(2) & 0.3478(5) & 0.3600 \\
 Fe-(Ch) height ({\AA}) & 1.50(1) & 1.505(7) & 1.446(3) & 1.434(3) & 1.315(5) & 1.46(1) \\
 Fe-(Ch) bond length ({\AA}) & 2.457(1) & 2.457(2) & 2.422(2) & 2.394(2) & 2.294(4) & 2.334(1) \\
 (Ch)-Fe-(Ch) bond angle (deg.) & 104.55$^{\rm b}$ & 104.44(1) & 106.70(5) & 106.37(8) & 110.0(1) & 107.1(1) \\
 Rp~($\%$) & -$^{\rm c}$ & R$_{1}$=5.89 & 21.3 & 18.0 & 24.7 & -$^{\rm a}$ \\
 Rwp~($\%$) & -$^{\rm c}$ & wRF$^{2}$=13.2 & 14.6 & 13.2 & 20.2 & -$^{\rm a}$ \\
 Chi$^{2}$ & -$^{\rm c}$ & 8.04 & 3.25 & 0.478 & 4.86 & -$^{\rm a}$ \\
\end{tabular}
\end{ruledtabular}
$^{\rm a}$ parameter fixed; $^{\rm b}$ calculated; $^{\rm c}$ not
given in the paper.

$^{\rm d}$ from neutron diffraction on single crystal in
\textit{I4/m} supercell; a-axis was divided by $\sqrt{5}$ for the
comparison; $^{\rm e}$ average value of total Fe1 (16i) and Fe2
(4d) site occupancies calculated taking into account the
multiplicity of both iron sites; $^{\rm f}$ average z-positions of
Se1 (4e) and Se2 (16i) calculated taking into account the
multiplicity of both selenium sites.

\end{table*}

\begin{figure}
\begin{center}
\includegraphics[width=\linewidth]{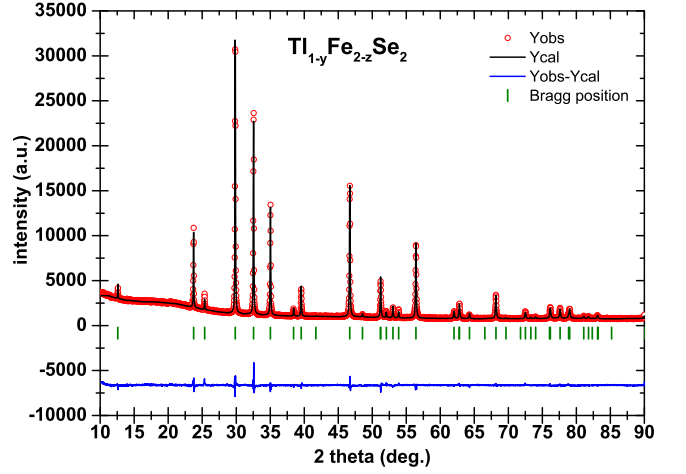}
\end{center}
\caption{Rietveld refinement profile of XRD pattern ($\lambda
$~=~1.5406~{\AA}) for Tl$_{1-y}$Fe$_{2-z}$Se$_{2}$ at room
temperature. A difference curve is plotted at the bottom (observed
minus calculated). Tick marks correspond to Bragg peaks of 122
selenide in the \textit{I4/mmm} space group description.}
\end{figure}

The structural parameters of
Tl$_{1-y}$Fe$_{2-z}$(Se$_{1-x}$S$_{x}$)$_{2}$ were refined from
XRD data by the Rietveld method using the ``Fullprof''
software~\cite{Rodriguez}. Data points with 10\r{}~$\le~2\theta~\le$~90\r{} were taken into account. A pseudo-Voigt profile shape
was used. The background was fitted using a linear interpolation
between selected points.

\begin{figure}
\begin{center}
\includegraphics[width=\linewidth]{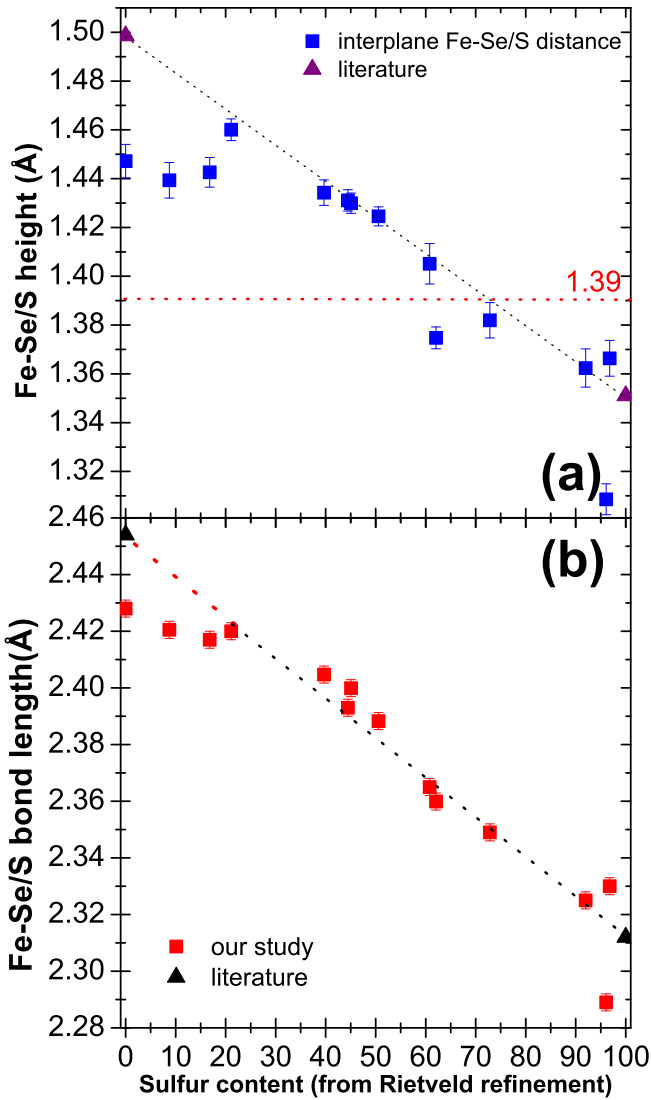}
\end{center}
\caption{(a) Fe-(Se,S) height (i.e. interplane Fe-(Se,S) distance) and (b) Fe-(Se,S) bond length of the Tl$_{1-y}$Fe$_{2-z}$(Se$_{1-x}$S$_{x}$)$_{2}$ samples versus the refined S content.}
\end{figure}

Attempts to use the space group corresponding to iron vacancy
ordering, for instance the tetragonal
$\sqrt{5}$~a~$\times~\sqrt{5}$~a~$\times$~c lattice, led to
refinement instabilities, due to the weakness of the
superstructure reflections. Consequently, all refinements were
carried out assuming the tetragonal \textit{I4/mmm} lattice (space
group No.139) i.e. neglecting ordering of iron vacancies.

The structure as described in the \textit{I4/mmm} space group
contains the following Wyckoff sites: Tl at 2a (0,0,0), Fe (in the
center of the FeCh$_{4}$ tetrahedron) at 4d (0,1/2,1/4) and Ch
(Ch=Se,S) at 4e (0,0,z) with z~$\sim0.355$ (with Se and S atoms
constrained to the same z coordinate). Refined variables were
lattice parameters, the z-position of the chalcogen atom, iron and
thallium occupancy factors and (Se,S) occupancy ratio (their total
summation was constrained to unity), and all isotropic Debye
Waller factors.

Figure 13 illustrates the result of the Rietveld refinement for
x=0 (pure selenide) as an example; there is a good agreement
between the experimental and the calculated profiles.

Table I. gives the refined structural parameters, bond lengths and
angles for x=0, x=0.5 and x=1. Our refined values for extremal x=0
and x=1 compositions are in good agreement with the old (see
table~I) and very recent literature~\cite{Haggstrom,Cao,Klepp}.

Figure 14(a) shows the evolution of the Fe-(Se,S) height, i.e. the
distance between the iron and the (Se,S) planes, with sulfur
content. As expected the substitution of Se by S with a smaller
atomic radius induces a continuous decrease of this inter-planar
distance (from 1.50~{\AA} for x=0 to 1.35~{\AA} for x=1); this
distance crosses the ideal value 1.39~{\AA} for which T$_{c}$ is
usually maximal in pnictides~\cite{Garbarino2} and the value
1.41-1.42~{\AA} where T$_{c}$ is maximal in FeSe under
pressure~\cite{Okabe}. In a similar way the Fe-(Se,S) bond length
decreases regularly with sulfur content in the lattice (see fig.
14(b)) from 2.45~{\AA} in Tl$_{1-y}$Fe$_{2-z}$Se$_{2}$ to 2.30~{\AA}
in Tl$_{1-y}$Fe$_{2-z}$S$_{2}$, i.e. covers the range explored
under high pressure on pure tetragonal FeSe (2.37~{\AA} at 0~GPa
to 2.29~{\AA} at 10~GPa) in which T$_{c}$ is
maximal~\cite{Garbarino1,Margadonna}. Nevertheless no
superconductivity is observed in our samples down to 4.2K.

\begin{figure}
\begin{center}
\includegraphics[width=\linewidth]{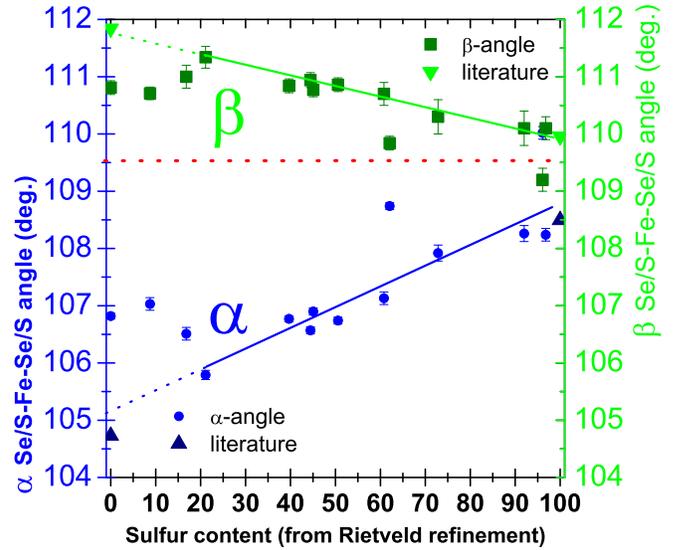}
\end{center}
\caption{(Se,S)-Fe-(Se,S) bond angle in the Fe(Se,S)$_{4}$
tetrahedron of the Tl$_{1-y}$Fe$_{2-z}$(Se$_{1-x}$S$_{x}$)$_{2}$ samples as a
function of refined S content.}
\end{figure}

Figure 15 shows the variation of the two characteristic
(Se,S)-Fe-(Se,S) bond angles in the Fe(Se,S)$_{4}$ tetrahedron
$\alpha$ and $\beta$ ($\alpha$ notation correspond to two (Se,S)
atoms of the same layer, while $\beta$ notation concerns two
(Se,S) atoms on both sides of the iron layer). The two
complementary angles merge towards the ideal value of the regular
tetrahedron (109.47~deg.), i.e. towards the value for which the
maximal T$_{c}$ is observed in superconducting iron-based
arsenides\cite{Lee}. Again, despite this, no superconductivity is found for
any (Se,S) composition.

It should be pointed out that the maximal T$_{c}$ value is not
achieved in FeSe iron selenide when the angle is ideal; on the
contrary T$_{c}$ values above 30K are reached for strongly
distorted tetrahedron~\cite{Garbarino1,Margadonna}.

\subsection{Discussion}

Rietveld refinements evidenced the continuous decrease of both
Fe-(Se,S) bond length and Fe-(Se,S) height with sulfur content in
Tl$_{1-y}$Fe$_{2-z}$(Se$_{1-x}$S$_{x}$)$_{2}$. As pointed out
above, despite the evolution of the structural parameters towards
values usually favoring superconductivity (i.e. optimal Fe-Ch
height and optimal Ch-Fe-Ch angle), superconductivity is not
induced by isovalent substitution of selenium by sulfur in this
iron-deficient Tl-122(Se,S) system. It seems that the only way to induce bulk superconductivty in the Tl-122 system is to replace partially Tl by an alkaline element, as observed by Fang et al. for potassium substitution, with T$_{c}$ around 30~K for samples rich in iron (z=0.18-0.22)~\cite{Fang}; they reported also superconductivity at T$_{c}$~$\sim$~20~K for Tl$_{1}$Fe$_{1.7}$Se$_{2}$ composition but with a very low superconducting volume fraction, suggesting a filamentary type superconductivity associated to a minority/impurity phase.

This difference of behavior could be related to slight structural
differences between pure thallium and alkaline element 122 systems. First
of all, the lattice volume, and then lattice parameters, of the
Tl-122 selenide are smaller than those of analogous compounds with
A~=~K,~Rb,~Cs by about $1-2\%$ (see table 1 in ref~\cite{Ye} for a
comparison). Secondly, the relative position of the chalcogen atom
relatively to the iron plane in the 122 selenides remains around the same value:
z=0.3530(2) in Tl-122(Se) (this work) compared to
z=0.3539(2)~\cite{Guo1}-0.3560(3)~\cite{Krzton} in K-122 and
z=0.3439(3)~\cite{Krzton}-0.3456(4)~\cite{Pomjakushin} in Cs-122.
This induces slightly shorter Fe-Se and Fe-Fe bond lengths in the
Tl-122(Se) compared to A-122 (A~=~K,~Rb,~Cs) by $1-2\%$. These differences
are enhanced when Se is substituted by S in Tl-122(Se,S), and this could
affect the electronic structure, and consequently the
insulating/superconducting behavior at low T.

The first DFT calculation was performed on stoichiometric
hypothetical TlFe$_{2}$Se$_{2}$ and revealed that the Fermi
surface is relatively close to the other iron-based compounds,
i.e. contains two electron cylinders, but with hole surfaces
suppressed~\cite{Zhang}. Electronic structure calculations were
then carried out on more realistic compositions: z=0.5 (with
orthorhombic $\sqrt{2}$ a $\times 2\sqrt{2}$ a $\times$ c
superstructure) and z=0.4 (with tetragonal $\sqrt{5}$ a $\times
\sqrt{5}$ a $\times$ c supercell) and compared with alkaline-based
analogous selenides~\cite{Yan,Cao1,Cao2}. In particular, it was
found that the Fermi surface of TlFe$_{1.6}$Se$_{2}$ is in fact
highly three-dimensional, unlike alkaline-based
selenides~\cite{Cao2}. Moreover, in the early calculations for z=0
(i.e. without iron vacancies), the density of states at the Fermi
level N(E$_{F}$) was found to decrease from
3.6-3.94~states/(eV.cell) in Cs or K intercalated
selenides~\cite{Nekrasov,Shein} to ca.
2~states/(eV.cell)~\cite{Zhang} in Tl-122(Se) selenide. On the other
hand, the full replacement of Se by S in K-122 was found to reduce
N(E$_{F}$) by ca. $50\%$ to 2.025~states/eV/cell for
KFe$_{2}$S$_{2}$~\cite{Shein}. This lower N(E$_{F}$) in the
Tl-based selenide, and even lower N(E$_{F}$) with S substitution,
could explain why x=0 and all S-substituted samples of the
Tl$_{1-y}$Fe$_{2-z}$(Se$_{1-x}$S$_{x}$)$_{2}$ series are not
superconducting at low temperature. This hypothesis has to be
checked theoretically (using realistic crystallographic structures
determined experimentally for the electronic structure
calculations) and experimentally. In that sense ARPES measurements
on pure thallium-based chalcogenides would be very useful.

Another important issue, still under debate, is the possible
existence of chemical/electronic phase separation at the nanoscale
in A$_{0.8}$Fe$_{2-z}$Se$_{2}$ selenides (A~=~K,~Rb,~Cs) suggested by
TEM structural studies~\cite{Wang}, synchrotron XRD~\cite{Ricci}
or STM studies of K-122 films~\cite{Li}. Very recently, based on
back-scattered electron images (SEM) and M\"{o}ssbauer
spectroscopy  Hu. et al. suggested that superconductivity of
K$_{0.8}$Fe$_{1.76}$Se$_{2}$ may be due to a sub-micron phase of
K$_{0.6}$Fe$_{1.9}$Se$_{2}$ composition~\cite{Hu}; Texier et al.
also reported a phase separation in a
Rb$_{0.74}$Fe$_{1.6}$Se$_{2}$ single crystal studied by NMR and
attributed superconductivity to the Rb$_{0.3(1)}$Fe$_{2}$Se$_{2}$
phase~\cite{Texier}. In the case of Tl-based 122 selenides, no
phase separation has been reported in the literature up to now; Tl
intercalated selenides appear more homogeneous, with a constant
iron content distribution and a nearly full Tl site, i.e. $y=0-0.1$
(contrary to A-122 selenides for which the alkaline site is more
deficient: $y=0.2-0.3$). All these results seem to imply that the doping
level (then the iron valency) is very different between the actual
superconducting A-122 phases and the Tl-122(Se,S) series, and this
could also explain why superconductivity is not observed in
Tl-122(Se,S) compounds.

\begin{figure}
\begin{center}
\includegraphics[width=\linewidth]{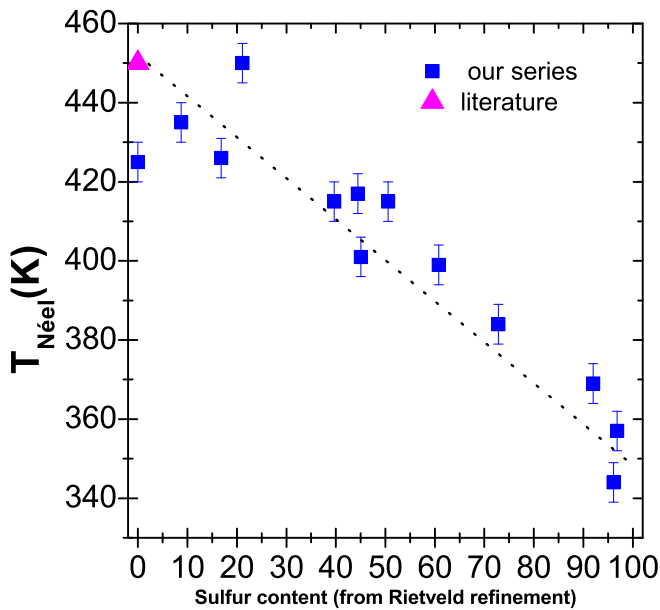}
\end{center}
\caption{N\'{e}el temperature (extracted from the electrical
resistance measurement) in
Tl$_{1-y}$Fe$_{2-z}$(Se$_{1-x}$S$_{x}$)$_{2}$ samples  versus the
refined S content.}
\end{figure}

Addressing now the high temperature magnetic behavior of the
Tl-122(Se,S) series, we have plotted N\'{e}el temperature values
(T$_{N}$) extracted from our transport measurements (see fig.~10 and 11)
vs the sulfur content x(S) in fig.~16. It shows a regular decrease
of T$_{N}$ with S content. As a consequence, there is a very good
correlation between T$_{N}$ and the Fe-(Se,S) height (see fig.
17): T$_{N}$ decreases continuously with the decrease of the
Fe-(Se,S) height. A similar trend is observed for a plot of
T$_{N}$ as a function of the Fe-(Se,S) bond length (not shown).

\begin{figure}
\begin{center}
\includegraphics[width=\linewidth]{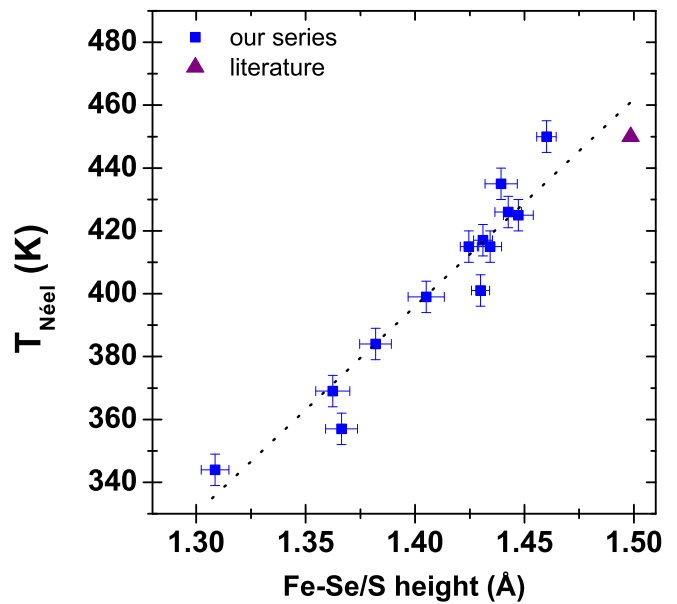}
\end{center}
\caption{N\'{e}el temperature in Tl$_{1-y}$Fe$_{2-z}$(Se$_{1-x}$S$_{x}$)$_{2}$ samples as a
function of the Fe-(Se,S) height in the structure.}
\end{figure}

We note that an opposite behavior has been reported in LaFeAsO
when As is substituted isoelectronically by Sb: T$_{N}$ decreases
also monotonously with antimony substitution, corresponding to an
increase of the equivalent Fe-(As,Sb) bond length~\cite{Karlsson}.
This behavior results from a complex competition between different
magnetic interactions in the system, i.e. the relative magnetic
exchange integrals between nearest iron neighbors and next-nearest
iron atoms in the same iron plane on one hand, and the magnetic
exchange between iron atoms from different planes on the other.
More theoretical work is necessary to interpret this linear
correlation with S content, determine the different magnetic
interaction energies and find the most stable magnetic
configuration. The related calculations were already made for the
$z=0.4$ and $z=0.5$ ($y=0$ in both cases) compositions~\cite{Cao2,Yan}.
It would be useful to extend it to sulfur-substituted
compositions. In addition it would also be interesting to study
the evolution magnetic excitations with sulfur content in the
thallium-based series and the differences with respect to their
superconducting alkaline-based analogues. In view of this, we
performed preliminary neutron diffraction experiments on
Tl-122(Se,S) to investigate their static long-range magnetic
structure. The results of this work will be published elsewhere.

\section{Conclusion}

The full solid solution of the
Tl$_{0.8}$Fe$_{1.5}$(Se$_{1-x}$S$_{x}$)$_{2}$ series (nominal composition), i.e. from
$x=0$ to $x=1$ was synthesized using the sealed tube technique. The
equivalent series with Se substituted by Te was also synthesized
up to $x(Te)=0.5$, but above $x(Te)=0.25$ samples were not monophasic.
The sulfur-based series was particularly studied by x-ray
diffraction, electron diffraction, magnetization and transport
measurements. No superconductivity was found down to 4.2K despite
that the optimal crystallographic parameters are reached in the
S-based series (i.e. the Fe-(Se,S) height and (Se,S)-Fe-(Se,S)
angle for which the critical superconducting transition T$_{c}$ is
usually maximal in pnictides). For Te-substituted samples we note
superconducting transitions, but probably related to the
tetragonal Fe(Te,Se) impurity phase. The S-based solid solution
shows a decrease of its N\'{e}el temperature (T$_{N}$), indicating
the onset of the antiferromagnetism order, from 450K in the
selenide ($x=0$) to 330K in the sulfide ($x=1$). Our structural
investigation emphasizes a direct linear relationship between
T$_{N}$ and the Fe-(Se,S) bond length.

\begin{acknowledgments}

The authors thank their collegues S. Karlsson, M.
N\'{u}\~{n}ez-Regueiro (Institut N\'{e}el) and G. Garbarino (ESRF)
for their support at the first stage of this work and for their
useful suggestions. We are also very grateful to Andr\'{e} Sulpice
(Institut N\'{e}el) for the low temperature magnetization
measurements of tellurium substituted samples and Yves Deschanels
(Institut N\'{e}el) for his help in the magnetization
measurements, in particular at high temperature. This work was
partially supported by the project SupraTetraFer ANR-09-BLAN-0211
of the Agence Nationale de la Recherche of France.

\end{acknowledgments}

\end{document}